\documentclass[preprint,12pt]{elsarticle}

\usepackage{amssymb}
\usepackage{amsmath}
\usepackage{multirow}
\usepackage{xcolor}
\usepackage{todonotes}

\journal{Nuclear Physics B}

\begin{document}

\begin{frontmatter}



\title{Quantifying Uncertainty in Void Swelling Prediction: A Conformal Prediction Framework for Reactor Safety Margins} 


\author{Minhee Kim}

\affiliation{organization={Department of Industrial and Systems Engineering, University of Florida},
            city={Gainesville},
            postcode={32603}, 
            state={FL},
            country={USA}}
\author{Yong Yang} 
\affiliation{organization={Department of Materials Science and Engineering, University of Florida},
            city={Gainesville},
            postcode={32603}, 
            state={FL},
            country={USA}}
\begin{abstract}
Irradiation-induced void swelling is a critical degradation mechanism for structural materials in nuclear reactors, dictating component operational lifespan and safety. While recent machine learning (ML) approaches have improved the accuracy of swelling rate predictions, they often fail to account for the inherent stochasticity of radiation damage, providing point estimates without rigorous uncertainty quantification. This lack of probabilistic context limits their applications in materials qualification, reactor licensing and risk assessment. In this work, we develop a framework that integrates ensemble ML models with Conformal Prediction (CP) to generate statistically calibrated prediction intervals. Unlike standard error estimation or Bayesian methods that often rely on rigid distributional assumptions, this approach specifically addresses the physical heteroscedasticity of swelling data, where variance transitions from the nucleation-dominated incubation regime to the growth-dominated steady-state regime. We demonstrate that log-transformed conformal prediction inference provides valid empirical coverage consistent with target confidence levels even in sparse data regimes. This framework offers a pathway to replace overly conservative upper-bound curves with Probabilistic Risk Assessment (PRA) tools for high-dose reactor core internals.
\end{abstract}

\begin{highlights}
\item A machine learning model is developed to predict void swelling behavior in irradiated materials from composition and irradiation conditions.
\item Conformal prediction enhances the model by calibrating uncertainty estimates with a prescribed coverage level on unseen experimental data.
\item Log-transformation improves robustness for heteroskedastic experimental data.
\item The model achieves a competitive predictive accuracy and offers empirical coverage consistent with the target confidence level.
\end{highlights}

\begin{keyword}
Void swelling \sep Machine learning \sep Uncertainty quantification \sep Conformal prediction \sep Error calibration



\end{keyword}

\end{frontmatter}



\section{Introduction}
\label{sec:introduction}

Irradiation-induced void swelling represents a critical microstructural degradation phenomenon for structural materials in high-radiation environments \cite{garner2006irradiation,mansur1978void,lucas1993evolution}. The progressive accumulation of voids leads to dimensional instability and significant degradation of mechanical performance, including embrittlement and fracture toughness reduction \cite{taller2021predicting, little1979void,neustroev2009severe, wang2016void}. Consequently, accurate prediction of swelling rates is essential for ensuring the structural integrity of core reactor internals and extending the operation license of nuclear energy systems \cite{sun2017influence}.

Traditionally, the prediction of void swelling has relied on a combination of empirical correlations derived from irradiation experiments and physics-based simulations (e.g. mean-field rate theory, phase-field modeling). Experimental studies provide the most reliable evidence, but are constrained by high costs and long lead times \cite{Jin2019a}. 
Consequently, most existing empirical studies have focused on characterizing how swelling behavior evolves in relation to a single variable at a time. Researchers have extensively explored individual factors, including displacement rate \cite{kalchenko2010prediction}, irradiation temperature \cite{yang2019influence}, the degree of cold-working \cite{wakai2002swelling}, and radiation type \cite{hudson1976void}. However, research examining how these variables simultaneously influence the swelling phenomenon remains limited, as fully factorial empirical experiments are often prohibitively labor-intensive. 

Conversely, physics-based models provide mechanistic insight but face significant challenges in bridging computational scales and accounting for the complex, coupled influence of multiple covariates simultaneously. For instance, Li et al. \cite{li2011phase} utilized a phase-field model to represent how thermodynamic and kinetic characteristics influence void formation and expansion in irradiated materials. Yet, these approaches often struggle to account for the combined influence of multivariate environmental factors. Furthermore, accurately simulating radiation damage by effectively bridging the gap between atomistic initiation and macroscopic swelling remains inherently challenging \cite{david2008study, tan2016microstructural}.

In recent years, Machine Learning (ML) has emerged as a powerful alternative, capable of mapping high-dimensional input features to material responses without requiring explicit functional forms \cite{mobarak2023scope}. 
Within the nuclear materials community, ML has been applied to a wide range of problems, including phase stability \cite{qin2024prediction}, mechanical properties \cite{stoll2021machine}, corrosion behavior \cite{allah2025application, hakimian2023application}, and irradiation response \cite{zhao2022application, sai2023machine}. For void swelling in particular, recent advancements have largely focused on three axes: maximizing predictive accuracy via AutoML and hybrid optimization \cite{Tang2026r, Pham2025f}, expanding databases through NLP and LLM integration \cite{Zhou2026l}, and enhancing interpretability using generative models like VAEs \cite{Liu2025d}. Notable successes include the prediction of incubation doses \cite{Jin2019a} and steady-state swelling rates \cite{Yang2024y}, with some studies significantly reducing prediction error for specific alloys \cite{yang2024prediction}.

\begin{figure}[t]
    \centering
    \includegraphics[width=1\linewidth]{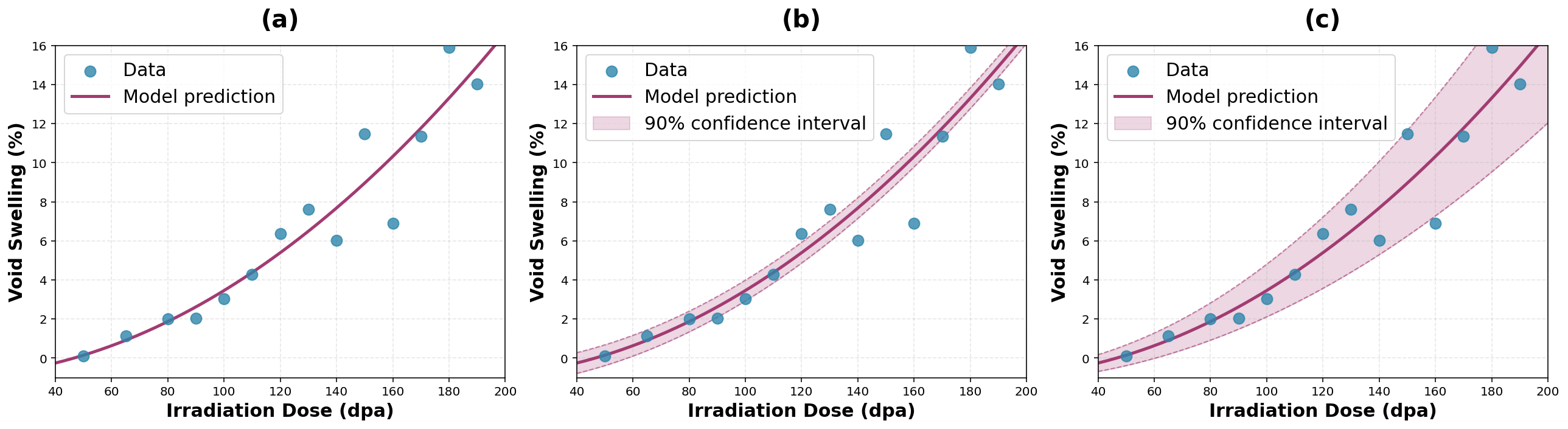}
    \caption{Illustration of UQ in predictive ML models: (a) Model without UQ, (b) Model with poorly calibrated UQ, and (c) Model with well calibrated UQ.}
    \label{fig:example}
\end{figure}

Despite these advances, most ML applications for void swelling remain deterministic, providing point predictions without a measure of confidence (uncertainty) (Figure \ref{fig:example}(a)). This limits their use in safety-critical applications where predictions inform regulatory decisions. While Bayesian methods \cite{Huh2024v, 9405460} have been explored for Uncertainty Quantification (UQ) in ML-based void swelling rate prediction, they often rely on specific distributional assumptions that may not accurately reflect the true swelling data. When these assumptions are violated, it can lead to severe miscalibration, where the nominal confidence interval fails to capture a significant fraction of observations (Figure \ref{fig:example}(b)). 

To overcome the aforementioned limitations, this work introduces a distribution-free UQ framework utilizing Conformal Prediction (CP). Unlike standard regression or Bayesian approaches that may struggle with the complex heteroscedasticity of irradiation swelling data, CP generates prediction intervals with rigorous mathematical coverage guarantees (e.g., 95\%) without assuming any specific error distribution or model structure \cite{Angelopoulos2021y, Shafer2008k}. This methodology effectively converts heuristic, uncalibrated estimates (Figure \ref{fig:example}(b)) into valid, statistically calibrated design corridors (Figure \ref{fig:example}(c)).
While CP has been successfully adopted in other high-stakes domains, such as prognostics \cite{Li2026FederatedConformal}, medical diagnosis \cite{vazquez2022conformal}, and robotics \cite{lindemann2023safe}, its application to void swelling has not been explored. 
In this study, a log-transformed CP inference layer is integrated with ensemble learning to demonstrate that rigorous safety margins can be maintained for void swelling predictions, even under sparse experimental data conditions.

\section{Methods}
\label{sec:methods}

To ensure rigorous evaluation and unbiased uncertainty calibration, the dataset is randomly partitioned into three disjoint subsets: a training set (80\%), a calibration set (10\%), and a test set (10\%). The training set is used to optimize the Random Forest regressor (Section \ref{ssec:ML}), while the calibration set is strictly reserved to compute non-conformity scores for CP (Section \ref{ssec:CP}). The final performance is evaluated on the held-out test set to assess generalizability to unseen irradiation conditions (Section \ref{sec:results}).

\subsection{Predictive machine learning model}
\label{ssec:ML}
We formulate the prediction of void swelling as a supervised regression problem. Let $x$ denote the vector of input features (e.g., irradiation dose, irradiation temperature, alloy composition), and let $y$ denote the corresponding measured void swelling rate. The objective is to learn the mapping $y=f(x)$ directly from data without imposing explicit functional forms.

To capture the complex and non-linear dependencies inherent to irradiation-induced void swelling, we employ a Random Forest (RF) regressor as the base predictive algorithm \cite{breiman2001random}. RF is an ensemble method that constructs $N$ independent decision trees during training. Each tree is fitted on a ``bootstrapped'' subset of the training data (sampling with replacement), and at each node split within a tree, only a random subset of features is considered. This combination of bootstrap aggregating (bagging) and feature randomness significantly reduces correlation among trees, making the final model more robust to outliers and overfitting than a single decision tree. The RF model was selected for this work due to its proven robustness against noisy experimental datasets and its successful application in prior ML studies of void swelling \cite{Jin2019a,Yang2024y}. 

For a given input vector $x$, the final ensemble prediction $\hat{y}(x)$ is calculated by averaging the local predictions $\hat{y}_t(x)$  of the individual trees $t$:
\begin{equation}
    \hat{y}(x)=\frac{1}{N}\sum_{t=1}^N \hat{y}_t(x).
\end{equation}

In addition to a point estimate, the RF provides a heuristic measure of predictive uncertainty. Since each tree represents a distinct realization of the model trained on a sub-dataset, the disagreement among trees serves as a proxy for model confidence. This heuristic uncertainty is calculated as the standard deviation of the predictions across the trees:
\begin{equation}
    \hat{\sigma}(x)=\sqrt{\frac{1}{N}\sum_{t=1}^N (\hat{y}_t(x)-\hat{y}(x))^2}.
\end{equation}
While this standard deviation indicates the internal variability of the model, it does not provide a formal statistical guarantee. Specifically, an interval defined by $\hat{y} \pm k\hat{\sigma}$ is not guaranteed to contain the true value $y$ with a specific probability. This limitation is especially crucial in void swelling applications, where data sparsity and increased variability at high doses may lead to underestimation of prediction error.

Standard regression models assume homoscedasticity, where the variance of the error term remains constant across the input space. A fundamental challenge in modeling void swelling is its inherent heteroscedasticity. In the low-dose incubation regime, swelling is governed by the stochastic nucleation of voids, heavily influenced by local solute trapping and precipitate interfaces, resulting in relatively contained data scatter. However, as materials transition into the steady-state swelling regime, the variance increases significantly. This ``explosion'' in variance at higher swelling rates is driven by complex interactions such as increased sink strength, saturation of dislocation structure, segregation-induced bias modification, and the coalescence of voids. A robust predictive model for nuclear materials must account for this dose-dependent expansion of uncertainty. 

Current deterministic ML models fail to capture this physical reality, potentially masking the risk of breakaway swelling events.
To address this, we implement a logarithmic transformation of the target variable:
\begin{equation}
    y_{log} = \ln(y + \text{offset}),
\end{equation}
where an offset of $1.0$ is applied to ensure all target values remain positive. The RF model is trained to predict $y_{log}$, and all subsequent predictions are mapped back to the original swelling scale through inverse transformation. This transformation ensures that the model remains sensitive to small swelling increments at low doses while stabilizing the variance for high-swelling outliers.

\subsection{Log-transformed conformal inference}
\label{ssec:CP}
A significant limitation of standard predictive ML models, including RF, is that they do not provide uncertainty estimates with a statistical coverage guarantee. In other words, a 90\% confidence interval obtained from the model’s internal variance is not mathematically guaranteed to contain 90\% of true observations. To bridge this gap, we employ CP, a general framework that calibrates uncertainty using observed prediction errors on a held-out dataset, thereby providing a formal coverage guarantee for new observations under broad conditions.

For each calibration sample $(x,y_{log})$, we compute a ``nonconformity'' score $s$, defined as the absolute prediction error in log space:
\begin{equation}
    s = \left| y_{log} - \hat{y}_{log}(x) \right|.
\end{equation}
These scores characterize the deviation of the ML predictions on data that were not used during the training phase. 
Given a target confidence level $1-\alpha$ (e.g., 90\%, 95\%), we calculate the calibration threshold $q_\alpha$ as the $(1-\alpha)$-quantile of these calibration scores $\{s\}$. This threshold effectively defines the width of the prediction interval.

For a new input $x^\star$, the conformal prediction interval in log space is constructed as:
\begin{equation}
    C_{log}(x^\star)=[\hat{y}_{log}(x^\star) - q_\alpha, \;\hat{y}_{log}(x^\star) + q_\alpha].
\end{equation}
Crucially, by construction, this interval is guaranteed to contain the true transformed response $y_{log}^\star$ with probability at least $1-\alpha$, subject only to the assumption of exchangeability between the calibration data ($\{(x,y)\}$) and test data ($\{(x^\star,y^\star)\}$):
\begin{equation}
\label{eq:coverage}
    P(y_{log}\in C_{log}(x^\star))\geq 1-\alpha.
\end{equation}

Formally, exchangeability requires that the calibration and test data are drawn from the same underlying data-generating process, such that their joint distribution is invariant under permutations. This assumption is a weaker and more flexible condition than the standard independent and identically distributed (i.i.d.) assumption widely used in most ML and statistical models.

Finally, the conformal interval is mapped back to the original physical swelling scale using the inverse logarithmic transformation:
\begin{equation}
    C(x^\star)=\left[
    \exp\left(\hat{y}_{log}(x^\star) - q_\alpha\right) - \text{offset},\;
    \exp\left(\hat{y}_{log}(x^\star) + q_\alpha\right) - \text{offset}
    \right].
\end{equation}
This yields an adaptive prediction interval for the void swelling rate that naturally widens at higher swelling values while remaining narrow near zero swelling, i.e., $P(y\in C(x^\star))\geq 1-\alpha$.

The key advantage of CP is that the coverage guarantee of the resulting prediction intervals in Equation \eqref{eq:coverage} is obtained directly from data without assuming any specific functional form or statistical distribution of the prediction errors. This makes the proposed approach robust to non-Gaussian noise and heteroscedasticity commonly observed in experimental void swelling measurements.

\section{Results}
\label{sec:results}
\subsection{Dataset and Preprocessing}
This study uses a comprehensive dataset of void swelling measurements compiled from published experimental literature and available in the Mendeley Data Repository (DOI: 10.17632/g4bnkxgc26).
The assembled dataset covers a broad range of irradiation temperatures, doses, irradiation types, and material compositions. A summary of the dataset features is provided in Table \ref{tab:datasummary}.

The predictive model utilizes an 18-dimensional input feature vector $x$ that consists of two categories of variables: irradiation variables and alloy compositional variables. The irradiation variables include irradiation dose (dpa), irradiation temperature (C$\circ$), irradiation gas environment (appm), and irradiation type. Alloy composition is represented by the weight percentages of 14 key elements: B, C, N, Al, Si, P, S, Ti, Cr, Mn, Fe, Ni, Cu, and Mo. These features collectively capture the synergistic effects of irradiation conditions and alloy compositions on swelling behavior. Among all features, only irradiation type is treated as a categorical variable, consisting of five distinct classes (Ni$^{6+}$ ion, Fe$^{2+}$ ion, Neutron, Proton, Electron).

The target response variable $y$ is the experimentally reported void swelling rate (\%). Inspection of the dataset reveals that a subset of records reports small negative swelling values, with a minimum value of approximately $-0.55\%$. Samples with negative swelling values were excluded from further analysis. After this exclusion, the final dataset contains 311 samples.

\begin{table}[t]
\centering
\begin{tabular}{cll}
\hline
Category & Name & Unit \\
\hline
\multirow{4}{*}{Irradiation condition ($x$)} 
 & Irradiation dose & dpa \\
 & Irradiation temperature & °C \\
 & Irradiation type & categorical \\
 & Gas concentration & appm \\
\hline
\multirow{14}{*}{Alloy composition ($x$)}
 & Boron (B) & wt.\% \\
 & Carbon (C) & wt.\% \\
 & Nitrogen (N) & wt.\% \\
 & Aluminum (Al) & wt.\% \\
 & Silicon (Si) & wt.\% \\
 & Phosphorus (P) & wt.\% \\
 & Sulfur (S) & wt.\% \\
 & Titanium (Ti) & wt.\% \\
 & Chromium (Cr) & wt.\% \\
 & Manganese (Mn) & wt.\% \\
 & Iron (Fe) & wt.\% \\
 & Nickel (Ni) & wt.\% \\
 & Copper (Cu) & wt.\% \\
 & Molybdenum (Mo) & wt.\% \\
\hline
Target variable ($y$) & Void swelling rate & \% \\
\hline
\end{tabular}
\caption{Summary of input variables and target variable used in this study.}
\label{tab:datasummary}
\end{table}

Non-predictive identifiers, such as steel names and record indices, were removed from the feature matrix. Unlike many other ML algorithms,  the tree-based RF architecture is invariant to monotonic scaling, and thus, input features were not normalized.  Finally, consistent with the validation strategy outlined in Section \ref{sec:methods}, the processed dataset was randomly partitioned into the training (80\%), calibration (10\%), and test (10\%) subsets.

\subsection{Exploratory data analysis}

Prior to model training, exploratory data analysis was conducted to characterize the statistical properties of the void swelling dataset. The analysis focused on evaluating the distribution of the target variable, the range of continuous input features, the representation of categorical irradiation conditions, and correlations among variables.

\begin{figure}[t!]
    \centering
    \includegraphics[width=0.6\linewidth]{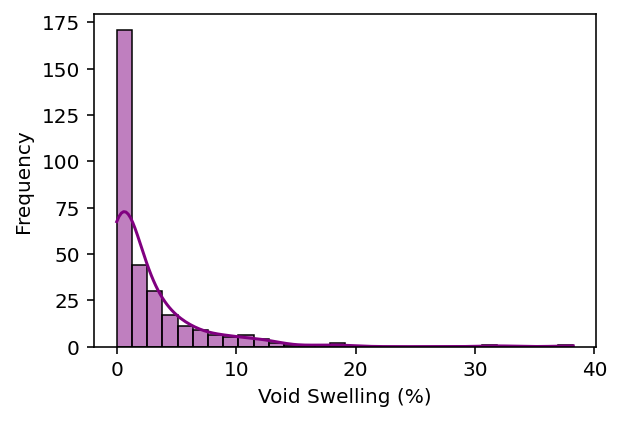}
    \caption{Distribution of void swelling (\%)}
    \label{fig:VSrateHist}
\end{figure}

Figure \ref{fig:VSrateHist} illustrates the distribution of the experimentally measured void swelling rates. The data is highly right-skewed, with the vast majority of measurements concentrated at low swelling values ($<5\%$) and a long tail extending to high-swelling regimes. A kernel density estimation (KDE) fit, overlaid on the histogram, provides a nonparametric representation of this underlying distribution and highlights its strong asymmetry. This skewness reflects typical experimental realities, where short-term, low-swelling observations are reported far more frequently than costly, long-term irradiations exhibiting high swelling rates.
Crucially, the extreme right-skewness of this target variable both motivates the logarithmic transformation and mathematically necessitates the distribution-free nature of the CP framework employed in this study.

\begin{table}[t]
\centering
\begin{tabular}{lccc}
\hline
Variable & Min & Mean & Max \\
\hline
Irradiation dose (dpa) & 0.50 & 80.96 & 228.76 \\
Irradiation temperature (°C) & 150.0 & 513.5 & 1023.4 \\
Gas concentration (appm) & 0.00 & 1.77 & 10.00 \\
B (wt.\%) & 0.000 & 0.0023 & 0.010 \\
C (wt.\%) & 0.000 & 0.051 & 0.110 \\
N (wt.\%) & 0.000 & 0.0097 & 0.150 \\
Al (wt.\%) & 0.000 & 0.034 & 1.120 \\
Si (wt.\%) & 0.000 & 0.462 & 1.270 \\
P (wt.\%) & 0.000 & 0.028 & 0.220 \\
S (wt.\%) & 0.000 & 0.0021 & 0.030 \\
Ti (wt.\%) & 0.000 & 0.163 & 1.110 \\
Cr (wt.\%) & 14.88 & 16.30 & 24.70 \\
Mn (wt.\%) & 0.000 & 1.41 & 1.94 \\
Fe (wt.\%) & 34.28 & 62.68 & 72.40 \\
Ni (wt.\%) & 8.40 & 16.92 & 43.00 \\
Cu (wt.\%) & 0.000 & 0.0044 & 0.540 \\
Mo (wt.\%) & 0.000 & 1.90 & 3.08 \\
\hline
\end{tabular}
\caption{Summary statistics of continuous input variables.}
\label{tab:summaryX}
\end{table}

As detailed in Table \ref{tab:summaryX}, the continuous input features ($x$) cover a broad parameter space. The irradiation conditions span wide operational ranges, particularly concerning dose and temperature. Simultaneously, the recorded alloy compositions cover both major alloying elements and minor trace constituents, with concentrations spanning multiple orders of magnitude.

\begin{figure}[thb]
    \centering
    \includegraphics[width=0.6\linewidth]{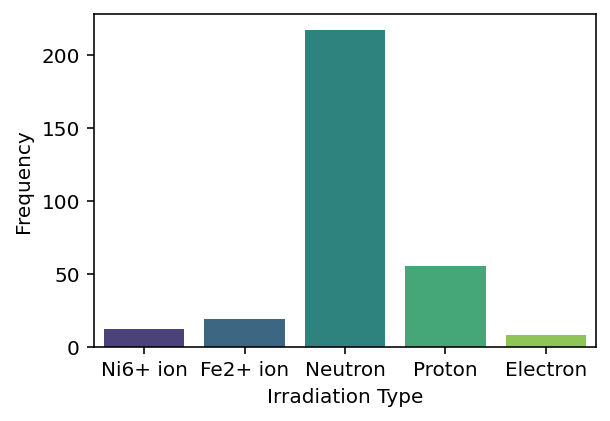}
    \caption{Distribution of irradiation types}
    \label{fig:irrtype}
\end{figure}

In addition to the continuous input features, the dataset includes one categorical variable, irradiation types. Figure \ref{fig:irrtype} shows the distribution of these irradiation type modalities across the compiled dataset. The data distribution is notably imbalanced. Neutron irradiation serves as the primary source of experimental data ($n=217$), followed by proton irradiation ($n=55$). Conversely, heavy ion (Fe$^{2+}$, Ni$^{6+}$) and electron irradiations constitute a much smaller fraction of the dataset.

\begin{figure}[thb]
    \centering
    \includegraphics[width=0.9\linewidth]{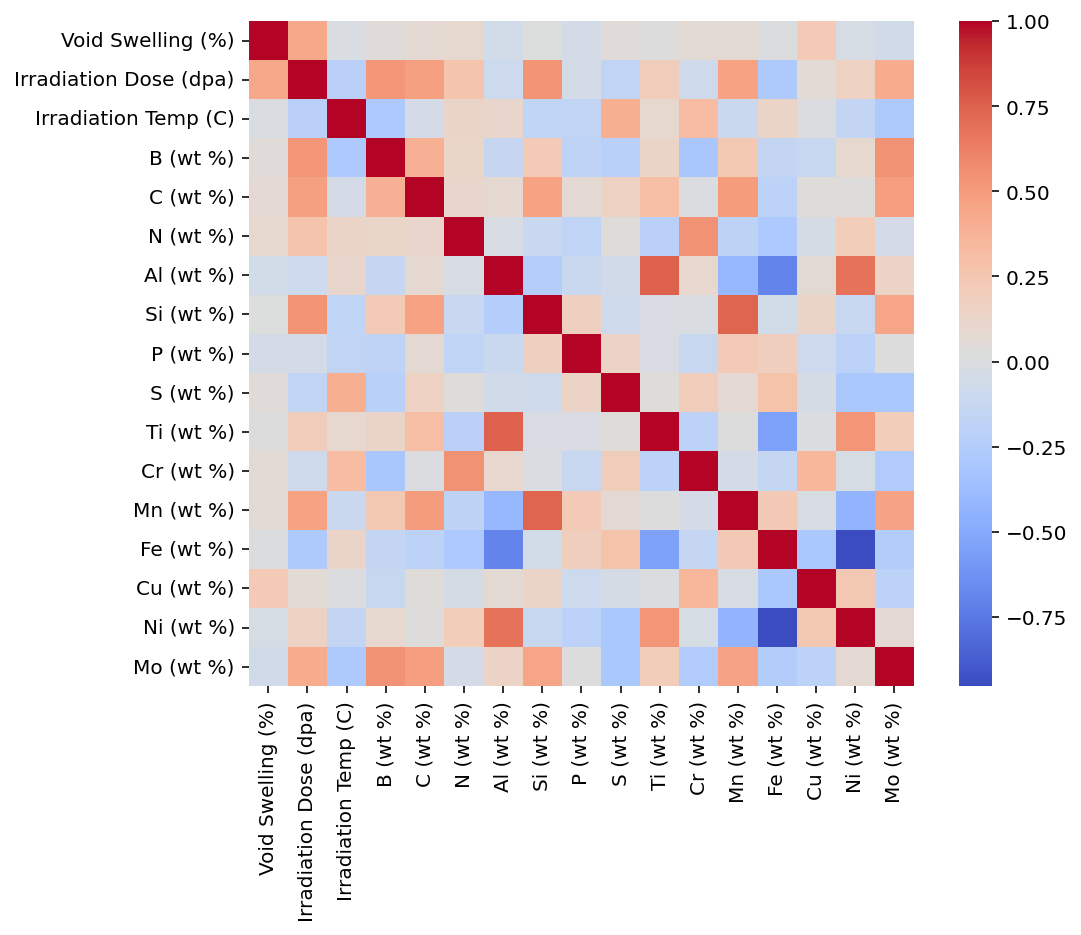}
    \caption{Correlation heatmap between continuous variables}
    \label{fig:heatmap}
\end{figure}

To assess linear interdependencies within the feature space, a Pearson correlation analysis was conducted (Figure \ref{fig:heatmap}). The results reveal a notable positive correlation between void swelling and irradiation dose, consistent with physical expectations. However, no strong linear correlations are observed between swelling and any other single input variables. This absence of simple pairwise linearity underscores that swelling behavior is governed by complex, multivariate synergies between chemical composition and irradiation conditions rather than independent factors, further validating the selection of a non-linear regression framework.

\subsection{Predictive accuracy}
The predictive accuracy of the proposed ML framework was initially assessed using a baseline RF model. Evaluated on the independent test set, the baseline model achieves a coefficient of determination ($R^2$) of $0.727$ and a Mean Absolute Error (MAE) of 1.82\%. While an $R^2$ above 0.7 indicates that a substantial fraction of the variance is captured, the remaining prediction errors motivate a closer examination of model performance.

\begin{figure}[tbh]
    \centering
    \includegraphics[width=0.5\linewidth]{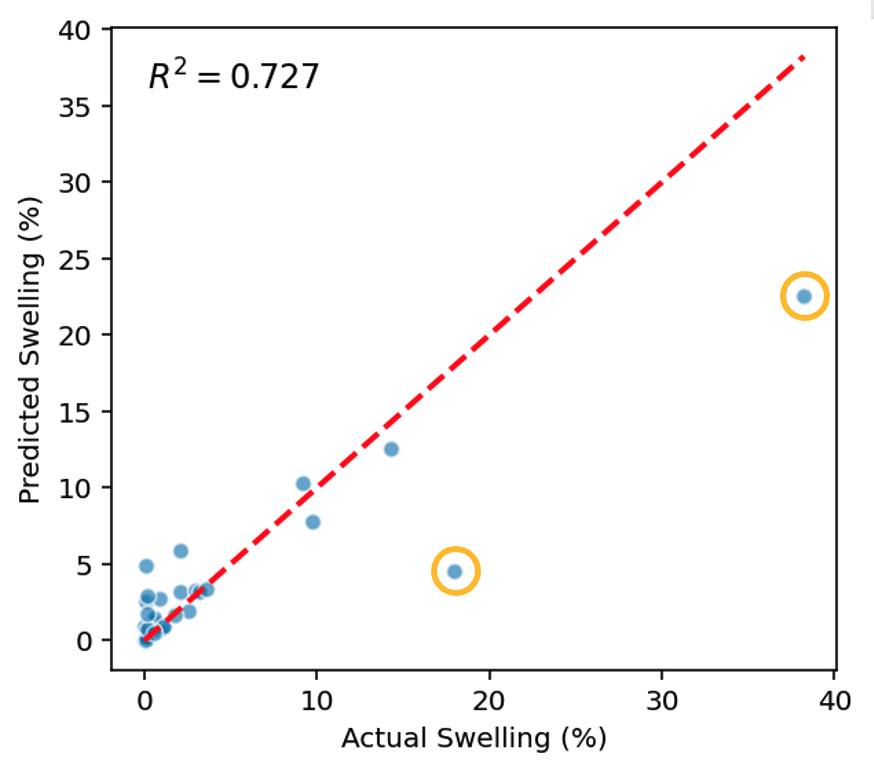}
    \caption{Parity plot of the RF model}
    \label{fig:parity}
\end{figure}

To understand the sources of prediction errors, we analyze individual test samples exhibiting large discrepancies between predicted and measured swelling values. The parity plot (Figure \ref{fig:parity}) reveals that while the model performs well in the low swelling rate regime, which dominates the dataset, prediction errors increase significantly at higher swelling rates (>15\%), due to data sparsity.
Two notable outliers (highlighted in Figure \ref{fig:parity}) illustrate this limitation. The first is a 316 SS irradiated with iron ions to an extreme dose of 220.56 dpa, where the model significantly underestimates the swelling (22.6\% predicted vs. 38.2\% measured). This underprediction likely stems from insufficient representations of extreme high dose swelling behavior in the training data. Similarly, the AISI 310 sample with high nickel (19.7\%) and chromium (24.7\%) contents exhibits a large positive residual (4.54\% predicted vs. 17.92\% measured). 

\begin{figure}
    \centering
    \includegraphics[width=\linewidth]{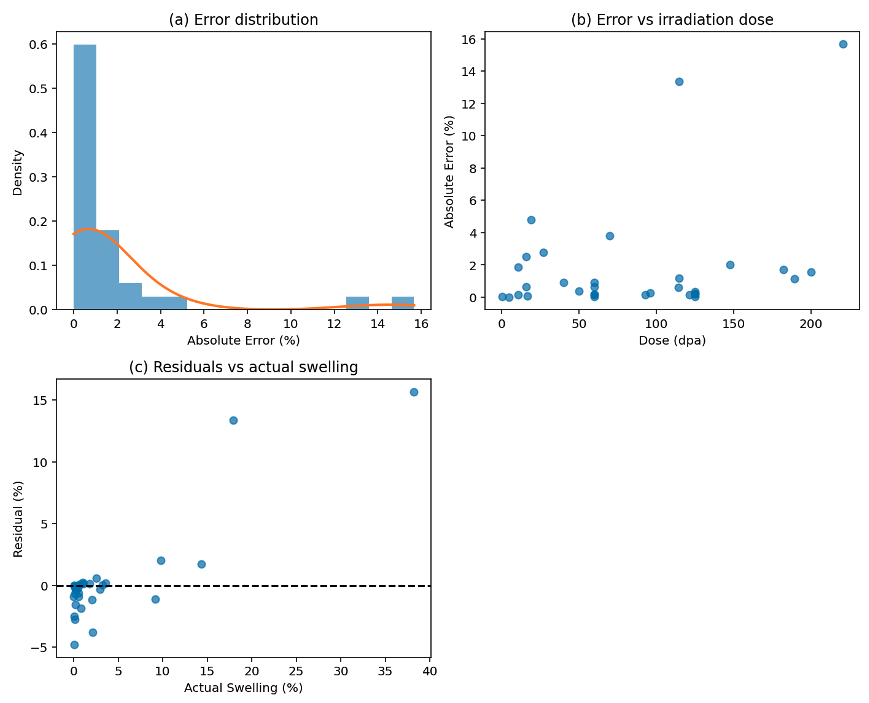}
    \caption{Error diagnostics for the RF model}
    \label{fig:errdist}
\end{figure}

This model instability is further evidenced by the distribution of absolute errors and residuals, which displays clear heteroscedasticity (Figure \ref{fig:errdist}). The error magnitude is not constant; rather, it increases significantly as a function of both the irradiation dose and actual swelling rate. Specifically, the absolute error remains relatively small for doses below 50 dpa, but residuals disperse rapidly beyond this threshold. 
These findings confirm that while the standard RF model provides a reasonable global point estimate, it becomes increasingly unreliable in high-swelling and high-dose regimes, precisely where the risk to structural integrity is greatest and data coverage is most limited. The presence of such outliers underscores the critical need for diagnostic tools and uncertainty-aware modeling approaches discussed in the following section. 

\subsection{Uncertainty quantification performance}
To address the limitations of deterministic point estimates, we evaluated the reliability and efficiency of the uncertainty intervals generated by the three distinct modeling approaches: the baseline Standard RF (without log transformation or CP), the Log-Transformed RF, and the proposed Log-Transformed CP framework. The quality of UQ was assessed using two primary criteria: uncertainty coverage and prediction interval width. 

Coverage quantifies the frequency with which the true experimental value falls within the predicted uncertainty bounds, directly measuring model reliability. For a target 80\% prediction interval, an empirical coverage significantly below 80\% indicates an overconfident and potentially unsafe model. Conversely, coverage substantially exceeding the target level suggests overly conservative bounds.

Interval width, defined as the difference between the upper and lower prediction limits, measures the precision and practical utility of the uncertainty estimate. While achieving high coverage is mathematically necessary, predicting impractically wide intervals renders the model useless. In the context of void swelling, tight uncertainty bounds are particularly crucial in the low-swelling regime to enable the early detection of swelling onset.

\begin{table}[h]
\centering
\begin{tabular}{lccc}
\hline
Model & Coverage (\%) & Avg. Width (\%) & Width ($<1$\%) \\
\hline
Standard RF & 68.75 & 2.93 & 2.38 \\
Log-transformed RF & 62.50 & 2.58 & 1.52 \\
\textbf{Log-transformed CP} & 84.38 & 6.86 & 3.46 \\
\hline
\end{tabular}
\caption{Comparison of uncertainty quantification performance for different models at a target confidence level of 80\%.}
\label{tab:coverage}
\end{table}

\begin{figure}[tbh]
    \centering
    \includegraphics[width=\linewidth]{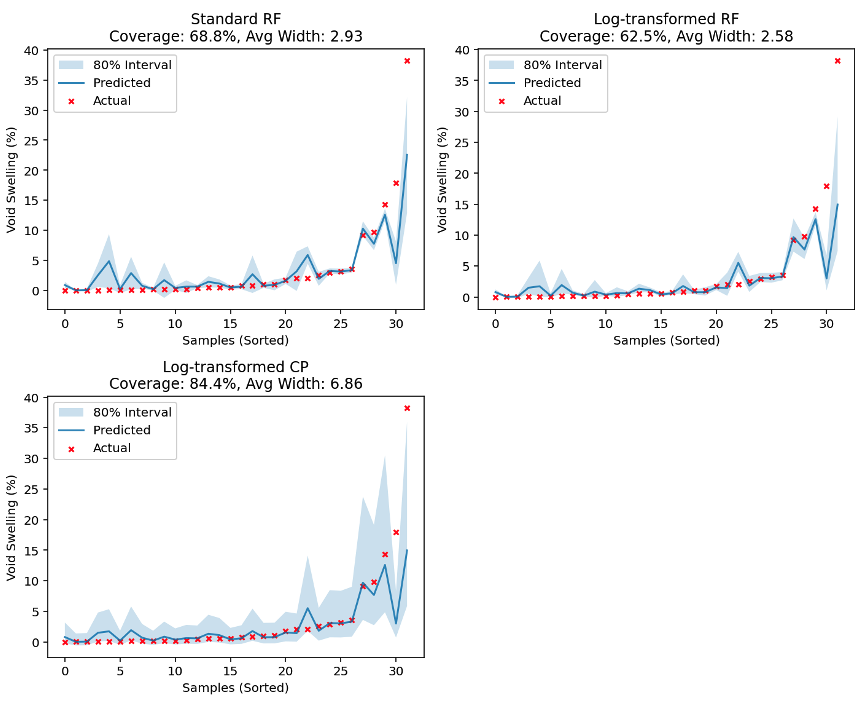}
    \caption{Void swelling rate prediction and associated 80\% uncertainty intervals for different models. Results are shown for test samples sorted by true swelling rate value.}
    \label{fig:UQ}
\end{figure}

Table \ref{tab:coverage} and Figure \ref{fig:UQ} summarize the UQ performance across the three models at a target confidence level of 80\%.
Both heuristic models (Standard RF and Log-transformed RF) failed to reach the coverage target, capturing only 68.75\% and 62.50\% of the test observations, respectively. This confirms that relying solely on the internal variance of the tree ensemble leads to severe overconfidence and underestimation of the true experimental errors.

In contrast, the proposed Log-transformed CP framework successfully met the statistical requirement, achieving an empirical coverage of 84.38\%. By utilizing a dedicated calibration set to calculate non-conformity scores, the CP model successfully calibrates the underlying ML baseline, ensuring statistically valid intervals for unseen data.

\begin{figure}
    \centering
    \includegraphics[width=0.65\linewidth]{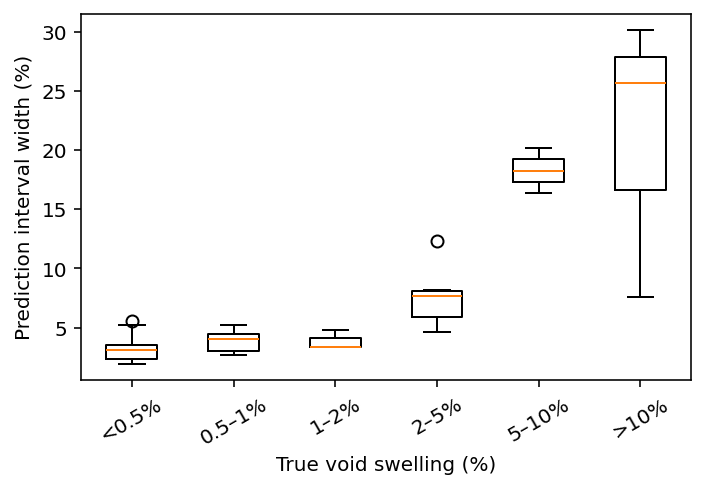}
    \caption{Boxplots of prediction interval widths of the proposed model grouped by true void swelling level.}
    \label{fig:boxwidth}
\end{figure}

This enhanced reliability naturally requires a trade-off in interval width. The average interval width for the Log-CP model was 6.86\%, which is notably more conservative than the nominally narrow but unreliable widths produced by the heuristic models (2.93\% and 2.58\%). Importantly, however, the Log-CP framework provides an adaptive interval. During the incubation period (swelling $<1\%$), the Log-CP model yields a narrower average width of 3.46\%. It expands appropriately in high-dose, high-uncertainty periods to maintain coverage guarantee, while narrowing significantly for low-swelling observations where the data is more precise. As demonstrated in Figure \ref{fig:boxwidth}, the distribution of prediction interval widths scales with true void swelling rates, effectively capturing the physical heteroscedasticity of the void swelling data.

\section{Discussion}
\label{sec:conclusion}

Standard ML models, despite achieving reasonable predictive accuracy for radiation-induced void swelling rates, consistently exhibit overconfidence. The heuristic uncertainty intervals derived directly from the baseline RF models are excessively narrow, failing to meet the 80\% target empirical coverage. For safety-critical reactor core internals, such uncalibrated and overconfident intervals are highly concerning, as they can lead to nonconservative regulatory assessments and engineering decision-making.

Exploratory data analyses confirm the strongly heteroscedastic nature of void swelling data. Model residuals increase significantly in regimes of higher swelling rates ($>20\%$) and elevated irradiation doses ($>70$ dpa), precisely where experimental data becomes sparse. Heuristic or fixed-width uncertainty estimates derived from standard ML models fail to capture this dose-dependent behavior, systematically underestimating uncertainty in high-swelling rate periods.

To overcome these limitations, a CP framework was implemented to mathematically calibrate uncertainty directly from observed prediction errors. By coupling CP with a logarithmic transformation of the swelling rate, the proposed approach successfully achieves empirical coverage consistent with the prescribed confidence levels while maintaining practically useful interval widths. This framework effectively converts heuristic uncertainty estimates into rigorously calibrated intervals, guaranteeing a specified coverage level on unseen data.

A primary advantage over conventional UQ approaches is that CP is distribution-free, meaning that the coverage guarantee remains valid regardless of the underlying complexity, noise, or non-linearity of the experimental data, making it particularly suitable for experimental materials datasets. Furthermore, CP functions as a post-processing step that can be applied to existing regression ML models without altering the underlying predictive methodology.

Compared to Bayesian hierarchical approaches previously applied to void swelling data, the CP framework offers several practical advantages. Specifically, it eliminates the need to specify prior distributions or make rigid assumptions regarding the underlying noise structure. Furthermore, the method is computationally efficient and can be universally applied to various ML regression models as a post-processing calibration layer. Ultimately, this approach produces rigorous uncertainty intervals with mathematical coverage guarantees.

\section{Conclusion}
This study addresses the critical challenge of quantifying uncertainty in data-driven Machine Learning (ML) models for radiation-induced void swelling. While ML has demonstrated high predictive accuracy, the safety-critical nature of nuclear reactor licensing demands rigorous reliability guarantees rather than simple point estimates. By integrating a Conformal Prediction (CP) framework with ensemble tree-based learning, this work establishes a robust methodology for generating statistically calibrated design corridors.

The key findings and contributions of this work are summarized as follows:
\begin{enumerate}
    \item \textbf{Valid Quantification of Stochasticity}: Unlike standard regression models that assume homoscedastic error, the log-transformed CP framework successfully captures the physical heteroscedasticity of void swelling. It correctly models the transition from the low-variance incubation regime to the high-variance steady-state growth regime, providing a valid coverage guarantee that is robust to the sparse and noisy nature of historical irradiation databases.
    \item \textbf{Distribution-Free Reliability}: A distinct advantage of this framework over recent Bayesian approaches is its independence from distributional assumptions. The framework produces valid prediction intervals without requiring the underlying swelling data to follow a specific distribution (e.g., Gaussian), making it particularly suitable for modeling the complex, non-linear behavior of breakaway swelling where normality assumptions often fail.
    \item \textbf{Transition to Probabilistic Design}: Practically, this framework enables a shift from deterministic, overly conservative upper-bound design curves to Probabilistic Risk Assessment (PRA). By providing a transparent ``trustworthiness'' metric, this tool allows reactor engineers and regulators to define specific risk tolerances (e.g., a 5\% probability of exceeding swelling limits), potentially extending the qualified lifespan of core internals and reducing unnecessary conservatism in component replacement schedules.
\end{enumerate}

Future work should focus on defining the precise domain of applicability, assessing transferability across material classes, and exploring active learning to target high-uncertainty regions in the irradiation parameter space.

\end{document}